\documentclass{iaus}
\usepackage{graphicx}

\title[Magnetic fields in Herbig stars] 
{Searching for a link between the magnetic nature and other observed properties of Herbig Ae/Be stars}

\author[S. Hubrig et al.]   
{S.~Hubrig$^1$,
 C.~Grady$^2$, M.~Sch\"oller$^3$, 0.~Sch\"utz$^1$, B. Stelzer$^4$, M.~Pogodin$^5$, M.~Cure$^6$ 
\and R.~Yudin$^5$}

\affiliation{$^1$ESO, Santiago, Chile;
email: shubrig@eso.org;
$^2$Eureka Scientific, Oakland, USA;
$^3$ESO, Garching, Germany;
$^4$INAF-Osservatorio Astronomico di Palermo, Italy;
$^5$Pulkovo Observatory, St.-Petersburg, Russia;
$^6$Univ. de Valparaiso, Chile }

\pubyear{2008}
\volume{259}  
\pagerange{1--2}
\date{?? and in revised form ??}
 \jname{Cosmic Magnetic Fields: from Planets,
to Stars and Galaxies } \editors{K.G. Strassmeier, A.G. Kosovichev
\& J. Beckmann, eds.}
\begin{document}

\maketitle

\begin{abstract}
We present the results of a new magnetic field survey of Herbig Ae/Be and A debris disk stars.
They are used to determine whether magnetic field properties in these stars are correlated
with the mass-accretion rate, disk inclinations,
companion(s), Silicates, PAHs, or show a
more general correlation with age and X-ray emission as expected for the decay of a remnant dynamo.

\keywords{stars: pre--main-sequence, stars: magnetic fields, X-rays: stars, techniques: polarimetric,
stars: individual (HD\,101412, HD\,139614, HD\,144668,  HD\,152404, and  HD\,190073)}
\end{abstract}

\firstsection 
\section{Introduction}
Protoplanetary disks are where planets form,
and where the pre-biotic materials, which  produce life-bearing 
worlds, are assembled or produced. We need to understand them,  how they 
interact with their central stars,  and their evolution;  both to reconstruct 
the Solar System's history, and to account for the observed diversity of exo-planetary systems. 
Our most detailed view of protoplanetary disks is for those surrounding intermediate-mass
stars, the Herbig Ae/Be stars (e.g. \cite{Herbig1960})
where the disks are revealed by their thermal
emission, and in scattered light in the optical and near-IR. 

\section{Observations and measurements}\label{sec:observ}

The observations were carried out in May 2008
at the European Southern Observatory with FORS\,1 
mounted on the 8\,m Kueyen telescope of the VLT.
New magnetic field detections were achieved in eight stars. For three Herbig Ae/Be stars,
we confirm the previous magnetic field detections. The star HD\,101412, with the largest magnetic field 
strength measured in our sample stars using hydrogen lines, $\left<B_{\rm z}\right>$\,=\,$-$454$\pm$42\,G ,
shows a change of the field strength by $\sim$100\,G during two consecutive nights.
In Fig.~1 (left panel) we present distinct Zeeman features detected at the positions of the 
hydrogen Balmer lines and the Ca~II H\&K lines.
The H$\beta$ line in the Stokes~$I$ spectrum is contaminated by the presence of 
a variable emission in the line core and was not included in our measurements.
Strong distinct Zeeman features at the positions of the Ca II H\&K lines are detected in four 
Herbig Ae/Be stars, HD\,139614, HD\,144668, HD\,152404, and HD\,190073. In Fig.~1 (right panel) we present 
the Stokes~$V$ spectra for these stars in the region around the Ca II doublet, together with our 
previous observation of HD\,190073. As we already reported in our earlier studies 
(\cite{Hubrig2004}; \cite{Hubrig2006}; \cite{Hubrig2007a})
these lines are very likely formed at the base of the stellar wind, as well as in 
the accretion gaseous flow, and frequently display multi-component complex structures in both the Stokes~$V$
and Stokes~$I$ spectra.
\begin{figure}
\centering
\includegraphics[height=1.8in,angle=0]{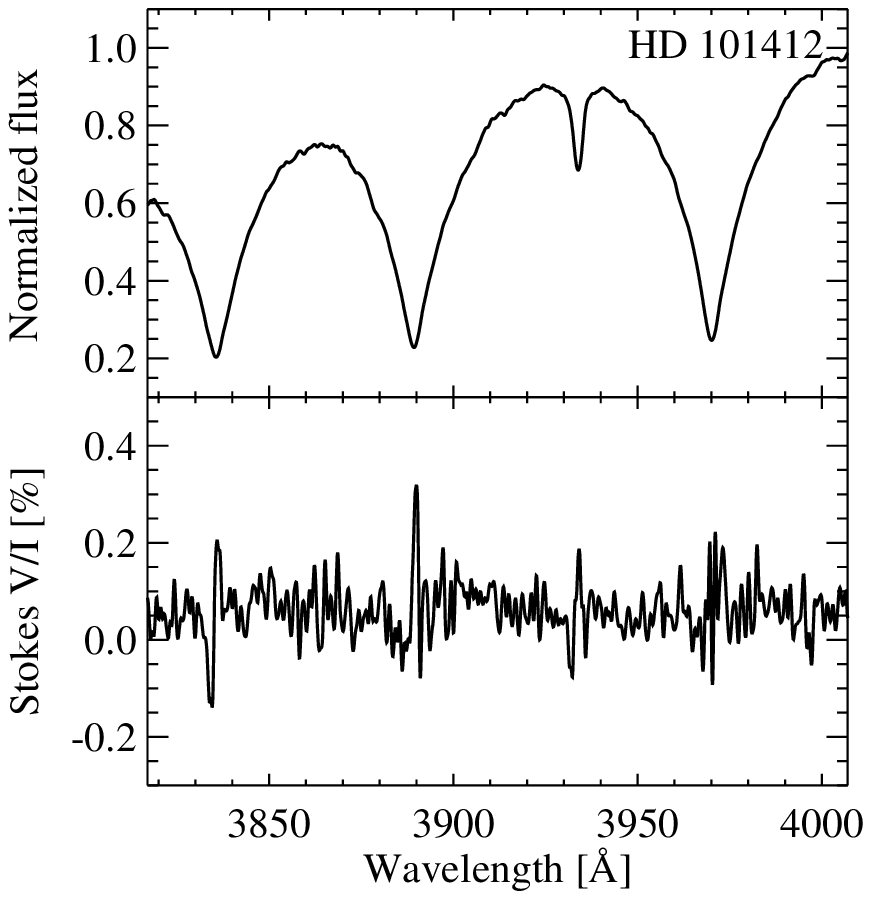}
  \includegraphics[height=1.8in,angle=0]{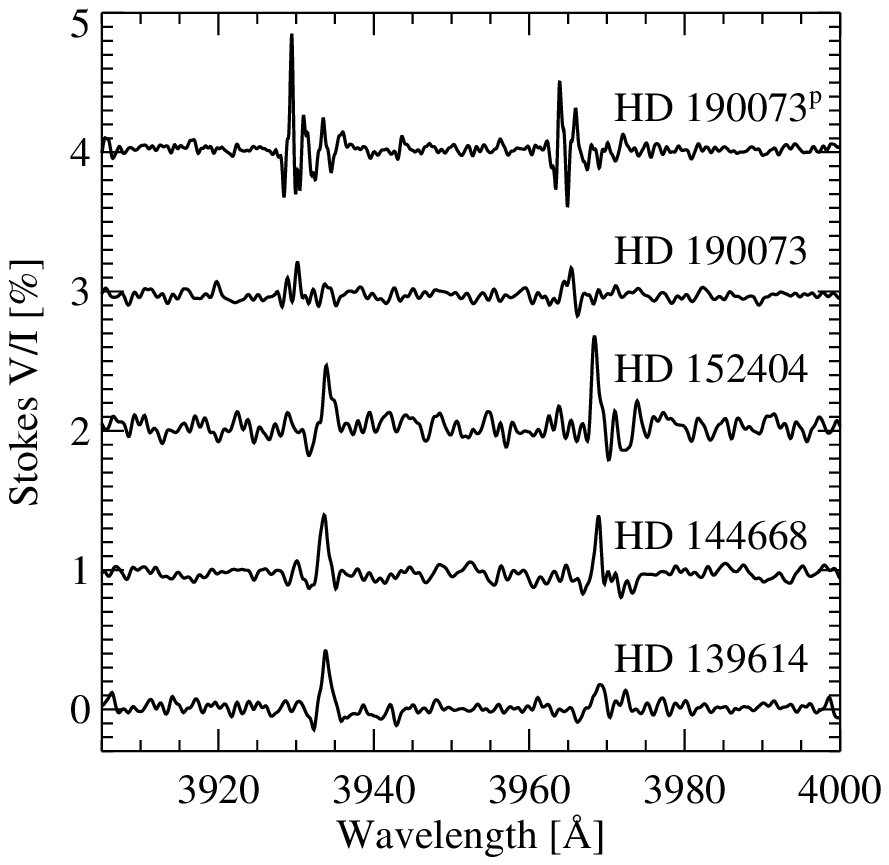}
 \caption{Left panel: Stokes~$I$ and $V$ spectra in HD\,101412;
Right panels: Stokes~$V$ spectra in the vicinity of the Ca~II H\&K lines of the Herbig Ae/Be stars 
HD\,139614, HD\,144668,  HD\,152404, and  HD\,190073. At the top we present the previous observation of
 HD\,190073, obtained in May 2005. The amplitude of the Zeeman features in the Ca~II H\&K lines 
observed in our recent measurement has decreased by $\sim$0.5\% compared to the previous 
observations.
 } \label{fig:Ca}
\end{figure}

\section{Looking for the links}
Most of our sample targets are Herbig Ae stars with masses of 3\,M$_\odot$ and less.
Since the observations of the 
disk properties of intermediate mass Herbig Ae stars suggest a close parallel to T\,Tauri stars, 
revealing the same size range of the disks, similar optical surface brightness and similar structure 
consisting of inner dark disk and a bright ring, 
it is quite possible that especially the magnetic fields 
play a crucial role in controlling accretion onto, and winds from, Herbig Ae stars, similar to the 
magnetospheric accretion observed in T\,Tauri stars.
We do not find any trend between the presence of a magnetic field, disk inclination angles and 
mass-accretion rates.
Also the membership in binary or multiple systems does not seem to have any impact on the 
presence of a magnetic field, whereas we find a hint that the appearance of magnetic fields is more 
frequent in Herbig stars with flared disks and hot, inner gas.
The stronger magnetic fields tend to be found in younger Herbig stars. The magnetic fields become very 
weak or completely disappear in stars when they arrive on the ZAMS, clearly confirming the conclusions of 
\cite{Hubrig2000} and \cite{Hubrig2007b} that magnetic Ap stars with masses less than 3\,M$_\odot$
are only rarely found close to the ZAMS.  
We also find a hint for an increase of the magnetic field strength with the level of the X-ray emission,
suggesting a dynamo mechanism to be responsible for the coronal activity in Herbig Ae stars.


%



\begin{thebibliography}{}

\bibitem[Herbig 1960]{Herbig1960}
Herbig, G.~H.\ 1960,
ApJS, 4, 337

\bibitem[Hubrig \etal{} (2000)]{Hubrig2000}
Hubrig, S., North, P., \& Mathys, G.\ 2000,
ApJ, 539,352

\bibitem[Hubrig \etal{} 2004]{Hubrig2004}
Hubrig, S., Sch{\"o}ller, M., \& Yudin, R.~V.\ 2004,
A\&A, 428, L1

\bibitem[Hubrig \etal{} 2006]{Hubrig2006}
Hubrig, S., Yudin, R.~V., Sch{\"o}ller, M., \& Pogodin, M.~A.\ 2006,
A\&A, 446, 1089

\bibitem[Hubrig \etal{} (2007b)]{Hubrig2007b}
Hubrig, S., North, P., \& Sch\"oller, M.\ 2007b,
AN, 328, 475

\bibitem[Hubrig \etal{} 2007a]{Hubrig2007a}
Hubrig, S., Pogodin, M.~A., Yudin, R.~V., et al.\ 2007a,
A\&A, 463, 1039

\end{thebibliography}
\end{document}